\documentclass[a4paper,11pt]{article}
\pdfoutput=1 

\usepackage{jinstpub} 
\usepackage{subfigure}
\notoctrue

\title{\boldmath Testbeam performance results \\ of bent ALPIDE Monolithic Active Pixel Sensors \\ in view of the ALICE Inner Tracking System 3}

\author[]{Bogdan-Mihail Blidaru on behalf of the ALICE collaboration}

\affiliation[]{Physikalisches Institut, Universit\"at Heidelberg, \\
Im Neuenheimer Feld 226, 69120 Heidelberg, Germany}

\emailAdd{m.blidaru@cern.ch}

\abstract{The ALICE Inner Tracking System has been recently upgraded to a full silicon detector consisting entirely of Monolithic Active Pixel Sensors, arranged in seven concentric layers around the LHC beam pipe. Further ahead, during the
LHC Long Shutdown 3, the ALICE collaboration is planning to replace the three innermost layers of this new ITS with a novel vertex detector. The proposed design features wafer-scale, ultra-thin, truly cylindrical MAPS. The new sensors will be thinned down to 20-40~{\textmu}m, featuring a material budget of less than 0.05\%~x/$\mathrm{X_{0}}$ per layer, unprecedented low, and will be arranged concentrically around the beam pipe, as close as 18~mm from the interaction point.

Anticipating the first prototypes in the new 65~nm CMOS technology node, an active R\&D programme is underway to test the response to bending of existing 50~{\textmu}m thick ALPIDE sensors. A number of such chips were successfully bent, even below the targeted innermost radius, without signs of mechanical damage, while retaining their full electrical functionality in laboratory tests. The curved detectors were subsequently tested during particle beam campaigns, where their particle detection performance was assessed.

In this contribution, testbeam highlights from the data analysis of bent ALPIDE sensors, will be presented. It was proved that the current ALPIDE produced in the 180~nm CMOS technology retains its properties after bending. The results show an inefficiency that is generally below $10^{-4}$, independent of the inclination and position of the impinging beam with respect to the sensor surface. This encouraging outcome proves that the use of curved MAPS is an exciting possibility for future silicon detector designs, as not only the sensor can survive the bending exercise mechanically, but the enticing attributes that make it attractive for use in the inner tracking layers are comparable to the flat state.}


\proceeding{12$^{\text{th}}$ International Conference on Position Sensitive Detectors - PSD12 \\
  12-17 September, 2021	\\
  Birmingham, U.K.}

\begin{document}
\maketitle
\flushbottom

\section{Roadmap to curved pixel sensors. The ITS3 detector concept}
\label{sec:its3}

ALICE is an experiment at the LHC that focuses on the study of heavy-ion collisions. 
It is designed to address the physics of strongly interacting matter at high center of mass energies. 
The detector ensemble is optimized for particle identification at low momenta and large particle multiplicities. 

A first Inner Tracking System (ITS) was successfully operated until the end of Run 2 at the very heart of ALICE, close to the interaction point.
Recently, the detector system was replaced by a second, digital-only version (ITS2), based on the ALPIDE Monolithic Active Pixel Sensors (MAPS). 
In this new design, a higher intrinsic spatial resolution 5~{\textmu}m is achieved \cite{felix_new}, owing to the high granularity of the 50~{\textmu}m thick sensors, the low material budget (0.35\%~x/$\mathrm{X_{0}}$ for the inner layers) and the proximity to the interaction point (23~mm for the innermost layer). 

\begin{table}[h]
\centering
\caption{\label{tab:params} Parameters of the ITS3 layout.}
\smallskip
\begin{tabular}{|l|ccc|}
\hline
Layer parameters                   & \multicolumn{1}{c|}{Layer 0}    & \multicolumn{1}{c|}{Layer 1}     & Layer 2  \\ \hline
Radial position (mm)               & \multicolumn{1}{c|}{18}         & \multicolumn{1}{c|}{24}          & 30       \\ \hline
Pixel sensor dimensions (mm)        & \multicolumn{1}{c|}{280 $\times$ 56.5} & \multicolumn{1}{c|}{280 $\times$  75.5} & 280 $\times$ 94 \\ \hline
Sensor thickness ({\textmu}m)              & \multicolumn{3}{c|}{20--40}                                                    \\ \hline
Pixel size ({\textmu}m$^{2}$) & \multicolumn{3}{c|}{$\mathcal{O}$ (10 $\times$ 10)}                                              \\ \hline
\end{tabular}
\end{table}

\begin{figure}[htbp]
\centering 
\includegraphics[width=0.78\textwidth]{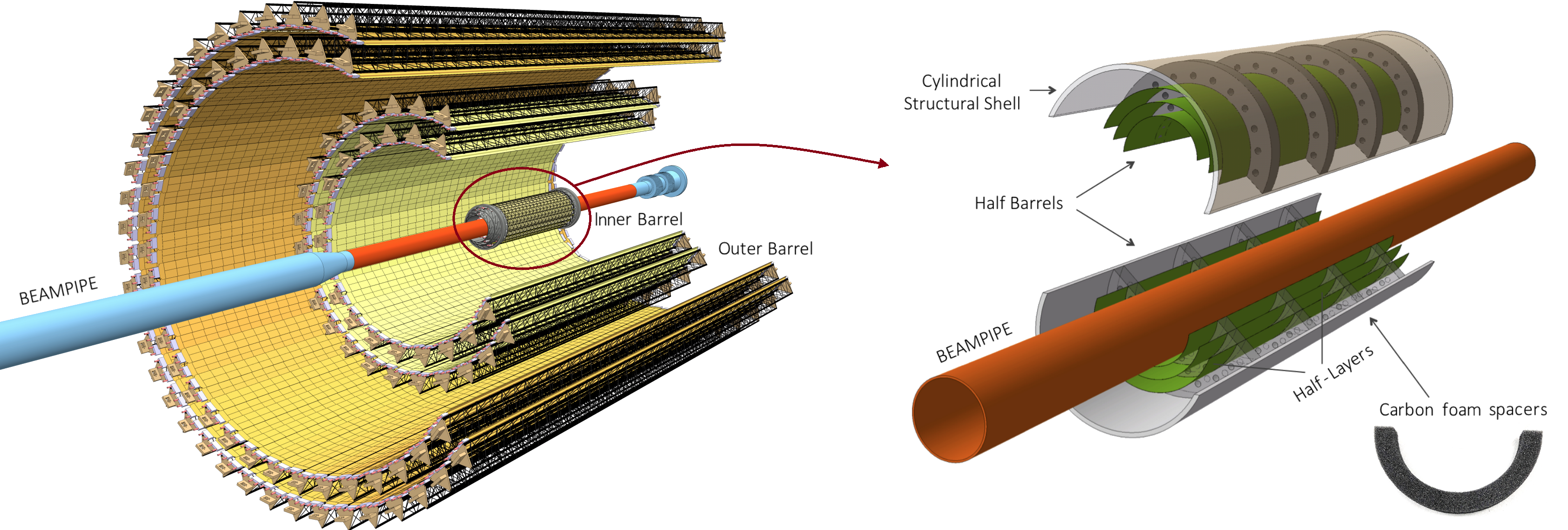}
\qquad
\caption{\label{fig:its2_and_its3}Left:~schematic of the seven concentric layers of the ITS2. The three layers of the Inner Barrel planned to be replaced are highlighted.  Right:~layout of the ITS3 Inner Barrel. The detector is made of two half-barrels, each containing three sensor layers. Ultra-light carbon foam spacers are used to define the radial position between the layers, as well as to offer mechanical support.}
\end{figure}

Even though the ITS2 detector is very lightweight, the active silicon part of the sensor represents only 15\% of the total material budget \cite{magnus_vertex2019_pos}. 
In order to fully reduce the material budget, the ALICE collaboration plans to replace the three innermost layers of the ITS2 with a novel, truly cylindrical, bent silicon tracker during the LHC Long Shutdown 3 (LS3), as shown in Figure \ref{fig:its2_and_its3}. 
The new CMOS sensors will be produced in the 65~nm technology node, as opposed to ALPIDE, which was fabricated in 180~nm technology. 
In this way it will be possible to access wafer-scale size sensors through stitching, a procedure provided by the foundry, where the pixel matrix of the chip is periodically repeated through aligned exposure of the mask. 
With these new MAPS, the material budget will be reduced to the minimum by removing the mechanical, electrical and cooling elements from the active area, thus reaching a value below 0.05\%~x/$\mathrm{X_{0}}$ per layer. 
Moreover, it is possible to place the first layer closer to the interaction point (18~mm), by replacing the beam pipe.

\section{Bending of ALPIDE sensors}
\label{sec:bending}

\begin{figure}[htbp]%
\centering
\subfigure[]{%
\label{fig:first}%
\includegraphics[height=1.15in]{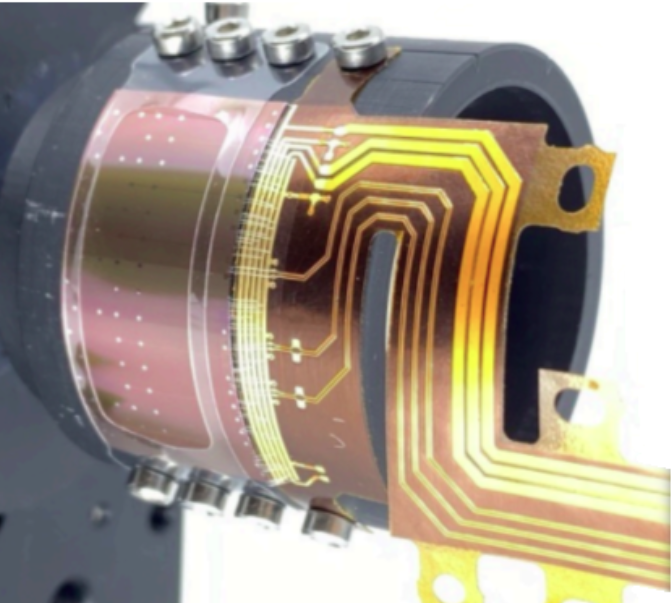}}%
\qquad
\subfigure[]{%
\label{fig:second}%
\includegraphics[height=1.15in]{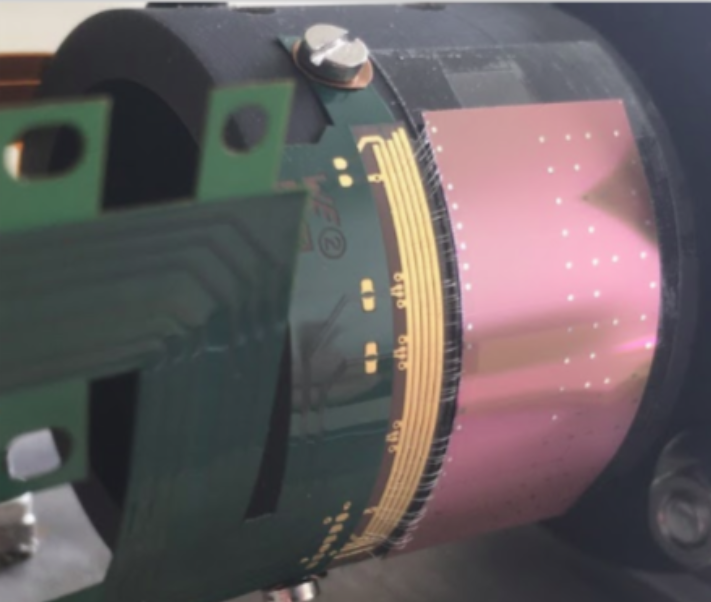}}%
\qquad
\subfigure[]{%
\label{fig:third}%
\includegraphics[height=1.15in]{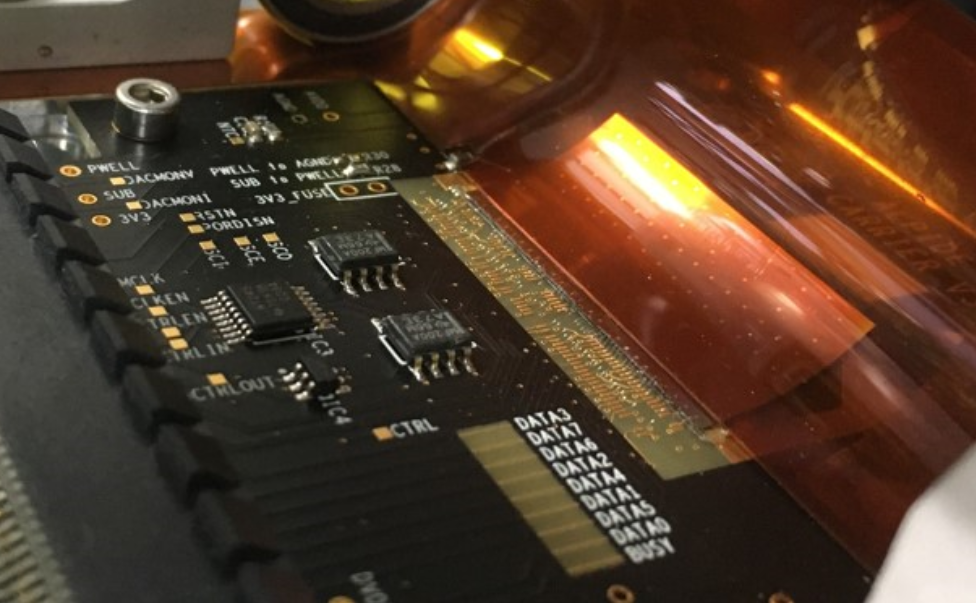}}%
\vspace*{-3mm}
\caption{\label{fig:bent_alpides}Different 50~{\textmu}m thick ALPIDE sensors bent to radii of 18~mm. (a) and (b) Sensors bent along the longer edge with different fastening mechanisms. (c) Sensor on carrier card bent along its short edge.}
\end{figure}

To study the feasibility of such a design, R\&D with 50~{\textmu}m thick ALPIDE sensors was started. Several such chips were bent in different configurations, as can be seen in Figure \ref{fig:bent_alpides} and to various radii, similar to the ones intended for the upgrade (see Table \ref{tab:params}).

\begin{figure}[htbp]
\centering 
\includegraphics[width=0.57\textwidth]{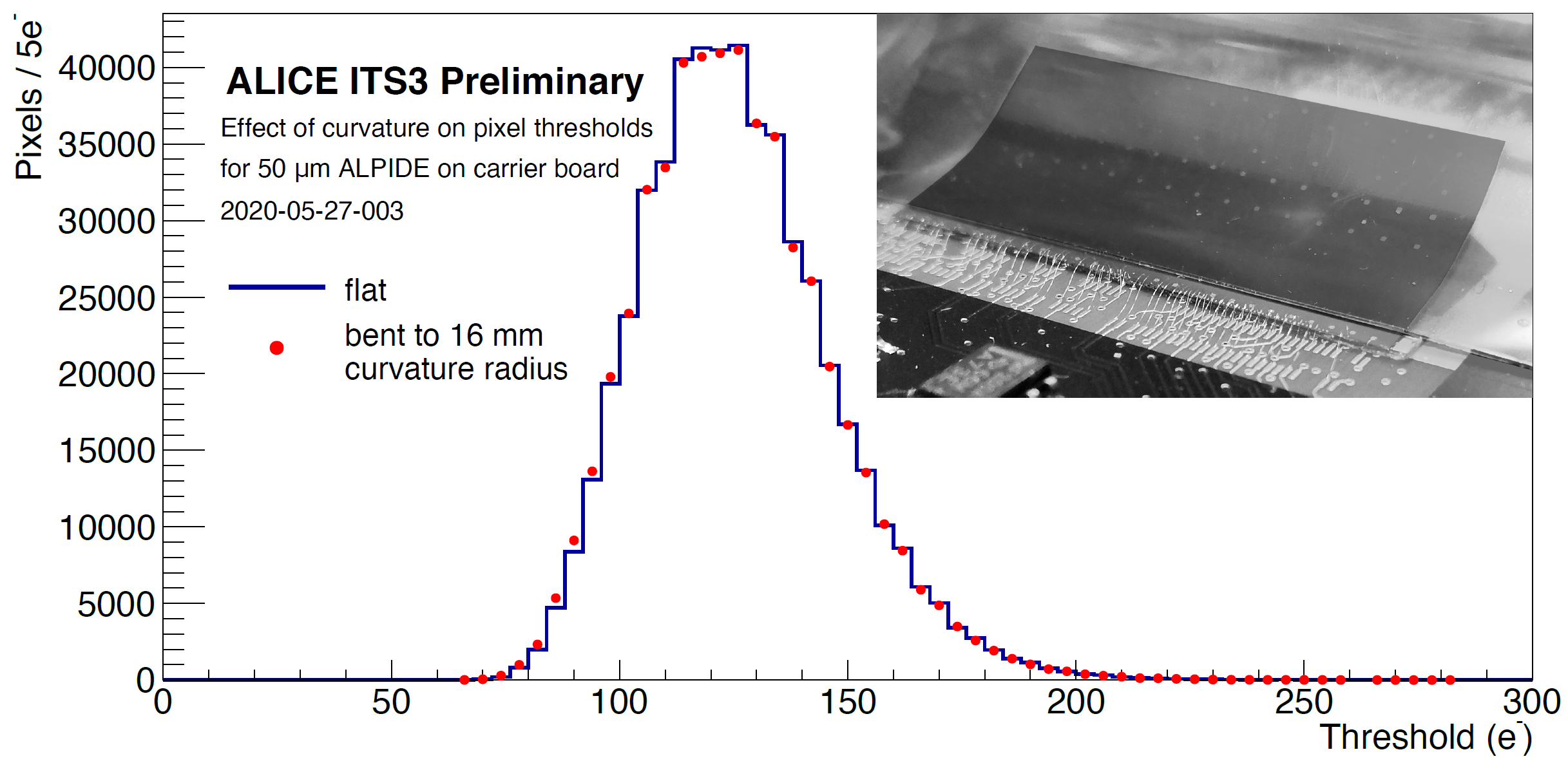}
\vspace*{-5mm}
\caption{\label{fig:threshold} Pixel-threshold distribution before and after bending the sensor to a radius of 16~mm \cite{first_paper}. Due to how it was fixed, the sensor relaxed in time, reaching a radius of 22~mm at the time of the beam test.}
\end{figure}

Their electrical performance and the mechanical stability was assessed, the sensors showing unmodified behavior with respect to their flat state in terms of threshold (see Figure \ref{fig:threshold}), noise levels and the number of faulty pixels. 

\begin{figure}[htbp]
\centering 
\includegraphics[width=0.99\textwidth]{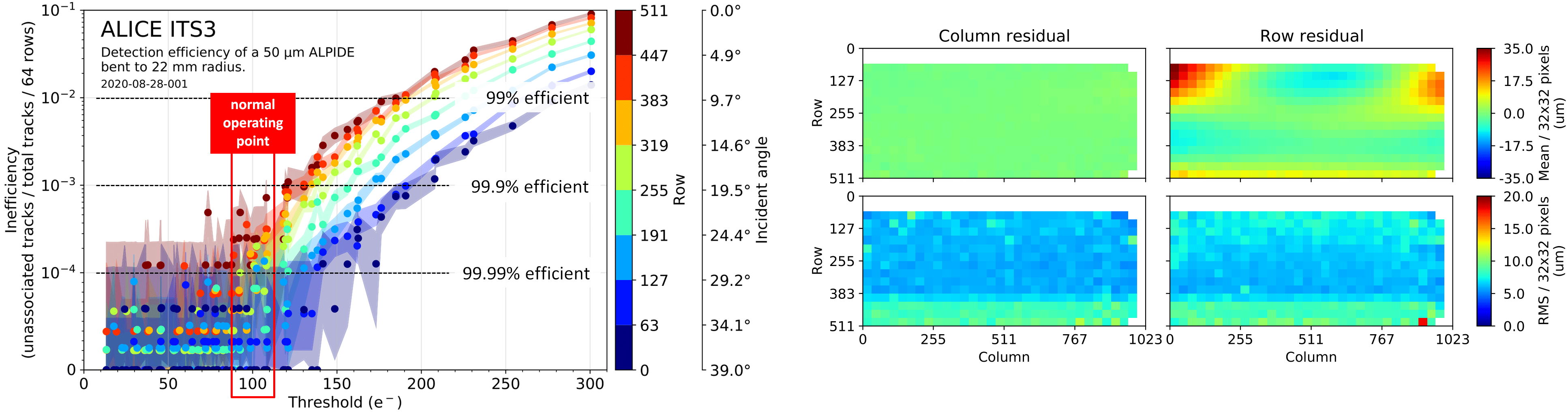}
\qquad
\caption{\label{fig:results} Left: hit detection inefficiency as a function of the chip threshold for different sensor rows, corresponding to various incident angles of incoming particles. Right: mean (top panels) and RMS (bottom panels) of the column and row spatial residuals \cite{first_paper}.}
\end{figure}

Following their initial assessment, the sensors were fully characterized \cite{first_paper} during a testbeam campaign at the DESY test beam facility with an electron beam of 5.4~GeV, using a telescope made of reference flat ALPIDE chips and treating the bent sensors as devices under test. In this document, results from the rightmost sensor shown in Figure \ref{fig:results} bent at a targeted radius of 22~mm are reported.

Data were taken by illuminating the whole sensor surface with the beam, for various device orientations and for a large range of pixel thresholds. The data were analyzed using the Corryvreckan \cite{corry} reconstruction framework.
The shape of the bent sensor was approximated by a purely cylindrical surface, with a radius given by surface metrology measurements.

In Figure \ref{fig:results} (left) the detection efficiency is shown as a function of the in-pixel sensor threshold for various beam incident angles. For the nominal threshold of ALPIDE (100~e$^{-}$) and below, the efficiency is larger than 99.9\%, independent of the incident angle or the position on the chip surface. Above the nominal threshold, the efficiency increases with increasing beam incident angles. This is explained by the larger amount of charge deposited by a particle traversing more material at a larger angle through the sensitive layer.

The right panel of Figure \ref{fig:results} shows the mean and the RMS of the spatial residuals in the column and row directions. The mean along the column direction is not influenced, since the sensor is invariant to rotations around the column axis. However, an effect is seen in the row direction, compatible with the cylindrical geometry and more pronounced towards the loosely attached corners of the sensor, leading to a slight sag in the middle of the sensor.
The RMS of the residuals shows a higher value towards large row numbers due to scattering where the chip is glued to the carrier card. For the row direction also an increase is visible towards small row numbers (high incident angles).

\section{Summary}

The ALICE collaboration plans to replace the Inner Barrel of the ITS2 during LHC LS3 with a new vertex detector featuring wafer scale, bent silicon pixel sensors.
In preparation for this big step, R\&D is progressing to study the effects of bending on MAPS. 
Recent testbeam measurements with curved ALPIDE sensors have proven the feasibility of bent MAPS. The devices were shown to retain their functionality and exhibit excellent performance after bending.

\acknowledgments

B.M.~Blidaru acknowledges support by the HighRR research training group [GRK~2058].

The measurements leading to these results have been performed at the Test Beam Facility at DESY Hamburg (Germany), a member of the Helmholtz Association (HGF). 



\begin{thebibliography}{99}

\bibitem{felix_new}
Felix Reidt, Upgrade of the ALICE ITS detector, arXiv:2111.08301 (2021).

\bibitem{magnus_vertex2019_pos}
Magnus Mager, Upgrade of the ALICE ITS in LS3, in \emph{Proceedings of The 28th International Workshop on Vertex Detectors {\textemdash} PoS(Vertex2019) {\textbf{373}} (2020)}.

\bibitem{first_paper}
The ALICE ITS Project, First demonstration of in-beam performance of bent Monolithic Active Pixel Sensors, arXiv:2105.13000v2 (2021).

\bibitem{corry}
D. Dannheim et al., Corryvreckan: a modular 4D track reconstruction and analysis software for test beam data, \emph{J. Instr. {\textbf{16}} (2021) P03008}.





\end{thebibliography}
\end{document}